\documentstyle[titlepage]{article}
\newcommand{\R}{{\rm I\kern-2pt R}}
\newtheorem{lemma}{Lemma}[section]
\newtheorem{proposition}{Proposition}[section]

\newtheorem{theorem}{Theorem}[section]
\newtheorem{remark}{Remark}[section]

\begin{document}
\begin{center}
{\bf Note on Purifications of a Qubit\\} 
Tiberiu Constantinescu and Viswanath Ramakrishna\footnote{Corresponding
 Author}\\
Department of Mathematical Sciences
and Center for Signals, Systems and Communications\\
University of Texas at Dallas\\
P. O. Box 830688\\
Richardson, TX 75083 USA\\
email: tiberiu, vish@utdallas.edu 
\end{center}

\begin{abstract}
This note provides an explicit parametrization of all purifications
of a mixed state in dimension 2 and all joint purifications, if any, of
two mixed states in the same dimension. The former is parametrized 
by $SO(3, R)$, while the latter is parametrized 
by $SO(2, R)$, except when the state being purified is already pure.
Using this, we show how to calculate certain
measures of quantum information and as a byproduct we show how to solve
one variation of the classical Procustes problem. 
This manuscript was originally scheduled to appear on the arXiv on
October 8th, 2002, but it did not due to alleged illegibility of 
the pdf version of the manuscript.   
\end{abstract}

\section{Introduction}
The notion of purification of a mixed state plays an important role 
in several contexts, \cite{nie}.
It provides insight into the question of decoherence.
It is important in quantum information theory from several points
of view. For instance, many quantitative measures such as the 
maximal singlet fraction may be explicitly defined in terms of
purifications.

The purpose of this note is to provide an explicit parametrization of
all possible purifications of a mixed state in two dimensions, and joint
purifications (if any) of two mixed states in a two dimensional
Hilbert space. In terms of this
explicit parametrization, this note recovers many of the
quantitative measures mentioned above. In particular, it shows that the
calculation of such measures reduces to optimization problems on 
the real orthogonal groups, $SO(2, R)$ and $SO(3, R)$.

The balance of this note is organized as follows. In the next 
section a precise definition of what we mean by purification and joint
purification is provided. The same section derives the parametrizations
referred to before. The III section shows how to reduce calaculations
of measures in quantum information to optimization problems, and gives
some instances of when this can be done in closed form. As a byproduct,
some insight into solving certain variations of the classical Procustes
problem is obtained. Section IV offers some conclusions.
In an attempt to extend this mixed states of higher
dimensions, we provide a Bloch sphere like characterization of three
dimensional mixed and pure states in the appendix. The final section
offers some conclusions.

\section{Parametrization of Purifications}

First, a mixed $2\times 2$ state is psd, trace $1$ matrix, $\rho$.
A purification of $\rho$ is a psd, trace $1$, projection, $P_{\rho}$, operating
in $C^{2}\otimes C^{2}$, such that the partial trace over the 
second $C^{2}$ factor of $P_{\rho}$ yields $\rho$.
Likewise, given a pair, $\rho_{1}, \rho_{2}$ of mixed states in $C^{2}$,
every pure state $P_{\rho_{1},\rho_{2}}$ in $C^{2}\otimes C^{2}$ whose
partial trace over the first (resp. second) $C^{2}$ factor is $\rho_{2}$
(resp. $\rho_{1}$) is said to be a joint purification of the pair
$\rho_{1}, \rho_{2}$.

Mixed states in $C^{2}$ may be represented in many fashions. However, one
suitable choice for the purpose at hand is the following
\[
\rho = \frac{1}{2} (I_{2} + \sum_{i=1}^{3}\beta_{i}\sigma_{i})
\]
where $\beta_{i}\in R, i=1, \ldots , 3$ and $i=1,2,3$ stands for
$i=x,y,z$ respectively. It is well known that $\mid\mid\beta \mid\mid \leq 1$,
with equality precisely for those $\rho$ which correspond to pure states.

For the same reasons, a mixed state in $C^{2}\otimes C^{2}$ are best 
represented in the following fashion
\begin{equation}
\label{intwod} 
\rho = \frac{1}{2} (I_{4} + \sum_{i=1}^{3}\beta_{i}\sigma_{i}\otimes I_{2}
+ \sum_{i=1}^{3}\gamma_{i}I_{2}\otimes\sigma_{i}
+\sum_{l=1}^{3}\sum_{k=1}^{3}\delta_{lk}\sigma_{l}\otimes \sigma_{k}) 
\end{equation} 
for real $\beta_{i}, \gamma_{i}, \delta_{lk}$. This representation is 
quite popular in the literature. However, to the best of our knowledge,
it is never refined further
to obtain a Bloch sphere like picture. This is needed for our purposes.
The following characterization of mixed states and pure states follows
from a direct calculation of squares of a Hermitian matrix 
(psd matrices are squares of Hermitian matrices and pure states are
psd matrices which equal their squares).
To avoid circumlocution, we call the vector of $\beta_{i}$'s as $\beta$, the
vector of $\gamma_{i}$'s as $\gamma$ and the
matrix of $\delta_{kl}$'s, $\delta$.

\begin{proposition}  
{\rm Every mixed state in $C^{2}\otimes C^{2}$ is of the form in Equation
(\ref{intwod}), with      
$\beta = \frac{1}{2}(\kappa\beta_{0} + \delta_{0}\gamma_{0})$, $\gamma =
= \frac{1}{2} (\kappa\gamma_{0} + \delta_{0}^{T}\beta_{0})$, $\delta
= \frac{1}{2}(\kappa\delta_{0} - ({\mbox adj} \ \delta_{0})^{T}
- \beta_{0}\gamma_{0}^{T})$, for any $\beta_{0}, \gamma_{0}\in
R^{3}$,  $\delta_{0}\in gl (3, R)$, $\kappa \in R$ satisfying
$\mid\mid \beta_{0}\mid\mid^{2} + \mid\mid\gamma_{0}\mid\mid^{2}
+ {\mbox Tr} (\delta_{0}^{T}\delta_{0})\leq 4$ and $\kappa$ is the
positive square root of the difference of the RHS and the LHS of this
inequality. Every pure state is of the form in Equation (\ref{intwod}), with
$\beta = \delta\gamma$, $\gamma = \delta^{T}\beta$, $\delta
= -[{\mbox adj} \ (\delta )]^{T} + \beta\gamma^{T}$, 
$\mid\mid \beta\mid\mid^{2} + \mid\mid\gamma\mid\mid^{2}
+ {\mbox Tr} (\delta^{T}\delta) = 3$.}  
\end{proposition}
 
Proof: The proof is a straightforward calculation. We will just record
one important calculation going into the verification of this proposition,
which will be needed for other purposes 
in this work.
If $\rho_{1}, \rho_{2}$ are two mixed states represented via the form
in Equation (\ref{intwod}), then the trace of their product is
\begin{equation}
\label{producttr} 
{\mbox Tr} \ (\rho_{1}\rho_{2}) = \frac{1}{4}(1 + <\beta_{1},\beta_{2}> 
+ <\gamma_{1},\gamma_{2}> + {\mbox Tr} \ (\delta_{2}^{T}\delta_{1}))
\end{equation}

Next, note that the
partial traces of such a $\rho$ (pure or impure) are precisely the
$2\times 2$ matrices $\frac{1}{2}(I_{2} + \sum_{i=1}^{3}\beta_{i}\sigma_{i})$,
$\frac{1}{2}(I_{2} + \sum_{i=1}^{3}\gamma_{i}\sigma_{i})$. This, of course,
directly implies that for any mixed state the lengths of $\beta$ and $\gamma$
is at most one. 
Further, a simple calculation,
left to the reader, shows 
i) ${\mbox det} (\delta ) = \mid\mid\beta\mid\mid^{2} - 1$ for a pure state;
ii) $\mid\mid\beta\mid\mid = \mid\mid\gamma\mid\mid$ for a pure state.

Returning to pure states, it follows from 
$\delta 
= -[{\mbox adj} \ (\delta )]^{T} + \beta\gamma^{T}$, 
$\gamma = \delta^{T}\beta$ and ${\mbox det} (\delta )
= \mid\mid\beta\mid\mid^{2} - 1$, that

\begin{equation}
\delta\delta^{T} = (1 - \mid\mid\beta\mid\mid^{2})
I_{3} + \beta\beta^{T} 
\end{equation}

The following result
says that this condition is essentially sufficient to determine
purifications of the state, $\frac{1}{2} (I_{2} + \sum_{i=1}^{3}\beta_{i}
\sigma_{i})$. This result also provides a complete parametrization of
such purifications.  

\begin{theorem}
{\rm Let $\frac{1}{2}(I_{2} + 
\sum_{i=1}^{3}\beta_{i}\sigma_{i})$ be a $2\times 2$
mixed state. 
Then all possible purifications, $P_{\rho}$ may be 
parametrized as matrices of the form in Equation
(\ref{intwod}), with $\beta$ the given the $\beta$, $\gamma =
\delta^{T}\beta$, $\delta$ any solution of the system of equations:   
\begin{eqnarray}
\label{eqnpurify} 
\delta\delta^{T} & = & (1 - \mid\mid\beta\mid\mid^{2})
I_{3} + \beta\beta^{T}\\ \nonumber
{\mbox \ det} \ (\delta ) 
& = &\mid\mid \beta \mid\mid ^{2} - 1 
\end{eqnarray}
This system of equations is always solvable. Further, the general solution
to this system (and thus the general purification, $P_{\rho}$) is provided
by $\delta = \tilde{\delta}S, S\in SO(3, R)$, with $\tilde{\delta}$ one
particular solution of this system, Equation (\ref{eqnpurify}).
Further, the set of purifications
of $\rho$ is parametrized by  
$SO(3, R)$ when $\mid\mid\beta\mid\mid < 1$ and by the unit
sphere, $S\in R^{3}$ when $\mid\mid\beta\mid\mid = 1$.}

\end{theorem}

{\it Proof:}
First, it is clear that any purification has to satisfy the
the system Equation (\ref{eqnpurify}). Proving the converse statement requires
first proving that the
respective system does have a solution, for any given $\beta$ within
the closed unit sphere in $R^{3}$,  and then that with
the choice of $\beta$, $\gamma$ and $\delta$ in the statement of 
the theorem, $P_{\rho}$ is indeed pure. In other words, this choice
of $\beta , \gamma , \delta$ indeed satisfies the  defining relations for
pure states, viz., $\mid\mid\beta\mid\mid^{2} +
\mid\mid\gamma\mid\mid^{2} + {\mbox Tr} (\delta^{T}\delta ) = 3$,
$\beta = \delta\gamma ; \gamma = \delta^{T}\beta $ and finally,
$\delta = - [{\mbox adj} \ (\delta )]^{T} + \beta\gamma^{T}$.

\noindent Case I: $\mid\mid \beta \mid\mid = 0$ . In this case $\beta$ is $0$.
So the system Equation (\ref{eqnpurify}) reduces to
\[
DD^{T} = I_{3}, \ {\mbox det}\ (D) = -1
\]
Clearly any matrix in $O(3, R)$ with determinant, $-1$, is a solution.
In this case, $\gamma = D^{T}\beta = 0$, and $D\gamma = D0 = 0 = \beta$.
So of the defining relations for pure states, only the first and the fourth
need checking. The first reduces to verifying ${\mbox Tr} \ [DD^{T}] = 3$
(since $\beta = \gamma = 0$), which obviously holds. The final equation now
becomes $D = -({\mbox adj} \ (D))^{T}$, which also holds since 
${\mbox det} \ (D) = -1$.

\noindent If $\tilde{D}\in O(3)$ is one solution to Equation (\ref{eqnpurify}),
then so is $\tilde{D}C , C\in SO(3, R)$. Conversely, if $D$ is a second
solution, then $C = \tilde{D}^{-1}D$ exists and satisifies,
\[
CC^{T} = \tilde{D}^{-1}DD^{T}\tilde{D^{T}}^{-1} = (\tilde{D^{T}}\tilde{D})^{-1}
= I_{3}
\]
Obviously ${\mbox det} \ (C) = 1$. So $C\in SO(3, R)$. 
   
\noindent Case II: $0 < \mid\mid\beta\mid\mid < 1$:
First, the matrix $(1 - \mid\mid\beta\mid\mid^{2})I_{3} +
\beta\beta^{T}$ is positive definite. Indeed $v^{T}
[(1 - \mid\mid\beta\mid\mid^{2})I_{3} + \beta\beta^{T}]v =
v^{T}v - \beta^{T}\beta v^{T}v + (<\beta , v>)^{2}$ is positive, for
$v\neq 0$, since $\mid\mid\beta\mid\mid < 1$. So there is always a matrix
$\delta$ satisfying $\delta\delta^{T} =
(1 - \mid\mid\beta\mid\mid^{2})I_{3} +
\beta\beta^{T}$.

Let us compute the determinant of such a $\delta$. We get
\[
({\mbox det}(\delta ))^{2} = (1 - \mid\mid\beta\mid\mid^{2})^{2} 
\]
Indeed, the eigenvectors of $(1 - \mid\mid\beta\mid\mid^{2})I_{3} +
\beta\beta^{T}$ are $\beta$ corresponding to eigenvalue $1$ and any two
vectors orthogonal to $\beta$ (in $R^{3}$) corresponding to the repeated
eigenvalue $1-\mid\mid\beta\mid\mid^{2}$. From this the previous equation
follows trivially. So, ${\mbox det} (\delta )$, for a given solution
$\delta$ of the first equation in the system 
Equation (\ref{eqnpurify}), is either the positive or negative
square root of $(1 - \mid\mid\beta\mid\mid^{2})^{2}$. To ensure the negative
square root, we multiply the given solution $\delta$ by $-I_{3}$ if needed. 
This also solves the first equation in Equation (\ref{eqnpurify}) and has the 
desired determinant.

Defining $\gamma = \delta^{T}\beta$, it is easy to verify  
$\beta = \delta \gamma $ and $\delta = -({\mbox adj}\ (\delta))^{T}
+ \beta\gamma^{T}$.
Indeed, $\delta\gamma = \delta\delta^{T}\beta
= (1 - \mid\mid\beta\mid\mid^{2})\beta + \beta\beta^{T}\beta = \beta$,
yielding the desired conclusion. To verify,  
$\mid\mid\beta\mid\mid^{2} +
\mid\mid\gamma\mid\mid^{2} + {\mbox Tr} (\delta\delta^{T}) = 3$, we note
first that upon taking trace on both sides of the first line in
the system (\ref{eqnpurify}) yields, ${\mbox Tr} \ (\delta\delta^{T})
= 3 - 2\mid\mid\beta\mid\mid^{2}$. Since ${\mbox Tr} \ (\delta^{T}
\delta ) = {\mbox Tr} \ (\delta\delta^{T})$, it suffices to show
that $\mid\mid\delta^{T}\beta\mid\mid^{2} = \mid\mid\beta\mid\mid^{2}$.
But $\mid\mid\delta^{T}\beta\mid\mid^{2} = \beta^{T}\delta\delta^{T}\beta
= \beta^{T} [(1 - \mid\mid\beta\mid\mid^{2})I_{3} + \beta\beta^{T}]\beta
= \mid\mid\beta\mid\mid^{2}$.
Finally, to show      
$\delta = - [{\mbox adj} \ (\delta )]^{T} + \beta\gamma^{T}$, we will
verify the transposed version.
Since, $\delta$ is invertible, we may premultiply both sides of
the first line in the system Equation ({\ref{eqnpurify}) by $\delta^{-1}$.
This, bearing in mind the second line of the system ({\ref{eqnpurify}),
yields
\[
\delta^{T} = - [{\mbox adj} \ (\delta )] + \delta^{-1}\beta\beta^{T}
= - [{\mbox adj} \ (\delta )] + \gamma\beta^{T}
\]
since it was just shown that $\delta\gamma = \beta$ holds.

To show that the $\delta$ of all purifications is given by    
$\tilde{\delta}C, C\in SO(3, R)$, with $\tilde {\delta}$ one particular
solution of the system (\ref{eqnpurify}), note first that $\delta
= \tilde{\delta}C$ trivially satisfies (\ref{eqnpurify}). Conversely,
if both $\delta$ and $\tilde{\delta}$ satisfy Equation (\ref{eqnpurify}),
the polar decomposition theorem plus 
the fact that both $\delta$ and $\tilde{\delta}$ have the same determinant
implies that there is a $C\in SO(3, R)$ such that $\delta = \tilde{\delta}C$
(for an argument which eschews the polar decomposition theorem see 
the remark following the proof).
  
Case III: $\mid\mid\beta\mid\mid = 1$ -
In this case the system (\ref{eqnpurify}) reduces to 
\begin{equation}
\label{lengthone}
DD^{T} = \beta\beta^{T}, {\mbox det} (D) = 0
\end{equation}.
Since per the first equation $DD^{T}$ is rank one, the second equation
is superfluous.
Once again there is at least one solution to Equation
(\ref{lengthone}), viz $\tilde{\delta } = \beta\beta^{T}$, for
the given $\beta$. 

Next, to verify that
any solution to Equation
(\ref{lengthone}), together with the given $\beta$ yields a purification,
we first observe that $\delta{\gamma} = \delta\delta^{T}\beta = \beta$,
since $\mid\mid\beta\mid\mid = 1$.
Just as in Case II, $\mid\mid\gamma\mid\mid = \mid\mid\beta\mid\mid = 1$.
This together with the obvious property that ${\mbox Tr}
\ (\delta^{T}\delta ) = \mid\mid\beta\mid\mid^{2} = 1$
yields $\mid\mid\beta\mid\mid^{2} + \mid\mid\gamma\mid\mid^{2} + {\mbox Tr}
\  (\delta^{T}\delta ) = 3$. To verify, the remaining condition, first
note that the matrix $\delta$ is also a rank one matrix.
Indeed, the rank of $\delta$ is the same as that of $\delta\delta^{T}$
(this is valid for any square matrix). So $\delta$ may be written in
the form $\delta = vu^{T}$ for some vectors $u, v \in R^{3}$.
So,
\[ 
{\mbox Tr} \ (\delta\delta^{T}) = {\mbox Tr} \ (vu^{T}uv^{T})
= \mid\mid u\mid\mid^{2}\mid\mid v\mid\mid^{2} =
{\mbox Tr}\  (\beta\beta^{T}) = \mid\mid\beta\mid\mid^{2} = 1
\] 
Hence, it holds that $\mid\mid u \mid\mid \  \mid\mid v\mid\mid =1$.
So, dividing $v, u$ by their lengths, if needed, 
it follows that $v$, $u$ may be chosen to be of length one.    
Now,\[
\delta\delta^{T} = vv^{T} = \beta\beta^{T}
\]
So either $v=\beta $ or  $v= -\beta$. Abosrbing the negative sign if needed
into $u$, we see $\delta = \beta u^{T}$ for a length one vector $u$.   
Since $u$ and $\beta$ have length $1$, and the 
group $SO(3, R)$ acts transitively on the sphere in $R^{3}$, 
it follows that there is some $C^{T}\in SO(3, R)$ such that
$u = C^{T}\beta$. So comparing $\tilde{\delta} = \beta\beta^{T}$ with
$\delta = \beta u^{T}$, we see $\delta = \tilde{\delta} C$.
Since $C\in SO(3, R)$, this verifies the claim.  

Finally, the assertion about the parametrization follows, since, when
$\mid\mid\beta\mid\mid < 1$, two purifications with 
distinct $\delta$ matrices are also distinct. When $\mid\mid\beta\mid\mid
= 1$, however, two purifications are distinct only if $C\beta\neq\beta$.
This means the redundancy in the parametrization consists of the isotropy
subgroup of $SO(3, R)$'s action at the point $\beta$, which
implies the stated condition on the parametrization
in this situation.    

\begin{remark}
Alternative argument for part of Case II: {\rm In the following a different
proof, which avoids the polar decomposition theorem, in Case II of the
previous proof is given. This calculation may be of interest in its own right.
Suppose $\delta$ and $\tilde{\delta}$ are two
solutions of Equation (\ref{eqnpurify}), then letting $C = 
(\tilde{\delta})^{-1}\delta $, it is clear that ${\mbox det}(C) = 1$.
Further,
\[
CC^{T} = (\tilde{\delta})^{-1} [(1-\mid\mid\beta\mid\mid^{2})I_{3}
+ \beta\beta^{T}] (\tilde{\delta}^{T})^{-1}
= (1-\mid\mid\beta\mid\mid^{2})[\tilde{\delta}^{T}\tilde{\delta}]^{-1}
+ \tilde{\gamma}\tilde{\gamma}^{T}, {\mbox where} \ \tilde{\gamma}
= \tilde{\delta}^{T}\beta
\]
A direct calculation shows that $\tilde{\delta}^{T}\tilde{\delta}
= (1-\mid\mid\beta\mid\mid^{2})I_{3} + \tilde{\gamma}\tilde{\gamma}^{T}$. 
Writing the matrix on the RHS of this last equation as $X + Y$,
we see that $X$ and $X + Y$ are both  
invertible and further $Y$ is of rank one. By a trivial modification of
the Sherman-Morrison-Woodbury formula, \cite{recipe},
it follows that $(X + Y)^{-1} = X^{-1} - \frac{1}{1 + {\mbox Tr} 
(YX^{-1}) }X^{-1}YX^{-1}$,
Applying this to $X =
(1-\mid\mid\beta\mid\mid^{2})I_{3}$, $Y = \tilde{\gamma}\tilde{\gamma}^{T}$,
yields
\[
CC^{T} = (1-\mid\mid\beta\mid\mid^{2})[\frac{1}{1-\mid\mid\beta\mid\mid^{2}}
I_{3} -
\frac{1}{1-\mid\mid\beta\mid\mid^{2}}\tilde{\gamma}\tilde{\gamma}^{T}]
+ \tilde{\gamma}\tilde{\gamma}^{T} = I_{3}
\]
i.e., $C\in SO(3, R)$.}
\end{remark}

\noindent {\it Joint Purifications:}
Next, we suppose that two mixed states are given, i.e.,
$\rho_{\beta} = \frac{1}{2} (I_{2} +\sum_{i=1}^{3}\beta_{i}\sigma_{i})$,
$\rho_{\gamma} = \frac{1}{2} (I_{2} +\sum_{i=1}^{3}\gamma_{i}\sigma_{i})$,
with {\it prespecified} $\beta , \gamma \in R^{3}$ are given. 
When does there exist a pure state $P$ in $C^{2}\otimes C^{2}$, 
such that the partial trace of $P$ over the second system yields $\rho_{\beta}$,
while that over the first yields $\rho_{\gamma}$.   
The aim is to parametrize all such $P$s.

Clearly, a necessary condition is that $\mid\mid\beta\mid\mid =
\mid\mid\gamma\mid\mid \leq 1$.
This is also sufficient.

\begin{theorem}
\label{jointpurify}
{\rm Suppose $\beta , \gamma \in R^{3}$ satisfy $\mid\mid\beta\mid\mid =
\mid\mid\gamma\mid\mid \leq 1$. Then they can be jointly purified.
Further, there is at least one 
solution, $\delta$, to the system Equation (\ref{eqnpurify})
which yields, per the prescription of Th 2.1, such a joint purification.
Given one such solution,
$\tilde{\delta}$, the most general joint purification is given by 
$\tilde{\delta}C$, with $C\in SO(3, R)$ satisfying $C\gamma = \gamma$.
This is a set parametrized by $SO(2, R)$, except when 
when $\beta =\gamma = 0$, in which case it is 
parametrized by $SO(3, R)$ or when 
$\mid\mid\beta\mid\mid = 1$, in which case there
is a unique joint purification.}     
\end{theorem}

{\it Proof:} Clearly, if we can find a solution $\delta$ to
the system Equation (\ref{eqnpurify}), which {\it further satisfies
the condition $\delta^{T}\beta = \gamma$ for the given $\beta , \gamma$},
the proof of Th 2.1 shows that the corresponding purification is 
indeed a joint purification. Suppose, for a specific solution,
$\delta_{sp}$, it holds that $\delta_{sp}^{T}\beta \neq \gamma$.
Then, computing
\[
\mid\mid\delta_{sp}\beta\mid\mid^{2} = \beta^{T}\delta_{sp}\delta_{sp}^{T} 
\beta = \beta^{T}\beta\beta^{T}\beta = \beta^{T} [1 -
\mid\mid\beta\mid\mid^{2}]\beta + \mid\mid\beta\mid\mid^{4}
= \mid\mid\beta\mid\mid^{2} = \mid\mid\gamma\mid\mid^{2}
\]
The last equation follows from the hypothesis $\mid\mid\beta\mid\mid
= \mid\mid\gamma\mid\mid$. Thus, $\delta_{sp}^{T}\beta$ has the same
length as $\gamma$. Now, $SO(3, R)$ acts transitively on spheres 
of any radius. So, there is a $C^{T}\in SO(3, R)$ satisfying the
condition $C^{T}\delta_{sp}^{T}\beta = \gamma$, i.e., $\tilde{\delta}
= \delta_{sp}C$ provides one joint purification. Now of the purifications
provided by $\delta = \tilde{\delta}C, C\in SO(3, R)$, 
only those purifications which satisfy $C^{T}\gamma = \gamma$ will 
yield a joint purification. Further, via the same arguments in Theorem 1,
the most general joint purification
is necessarily supplied by $C\tilde{\delta}$ with $C\in SO(3, R)$ 
such that $C\gamma = \gamma$. The collection of all such $C$'s is, of course,
the isotropy group at $\gamma$ of $SO(3, R)$'s action on this sphere.
This isotropy group is conjugate to the isotropy at the vector
$\mid\mid\gamma\mid\mid (1, 0, 0)$, which is precisely $SO(2, R)$.
Geometrically, all such $C$'s are rotations in the plane perpendicular to
the vector $\gamma$, while $\gamma$ is the axis of rotation.   
If $\mid\mid\beta\mid\mid < 1$, then due to the invertibility
of the $\delta$ matrix of purifications, it follows that $SO(2, R)$
paramterizes the collection of joint purifications.
If $\mid\mid\beta\mid\mid = 1$, then $\beta\gamma^{T} =
\beta (C\gamma )^{T}$, for every $C\in SO(3, R)$ fixing $\gamma$. So there
is just one joint purification.
Finally, if $\gamma = 0$, then the condition $C\gamma = \gamma$ is
no constraint on $C\in SO(3, R)$.

\begin{remark}
{\rm  It is, of course, possible to induce on the set of purifications
the additional structures in the orthogonal groups (or the sphere when
$\mid\mid\beta\mid \mid = 1$).
However, this may not be very useful.
For instance, the Riemannian metric on the orthogonal groups may not be
consistent with the any of the current notions of distance between 
pure states. However, in a certain sense, these additional structures will
be employed later in this work. More precisely, in the next section,
some calculations of quantum information measures will be reduced
to optimization on the orthogonal groups. The fact that these 
problems have a solution follows from the compactness of these groups.
Further, they can be reduced to optimization problems over products
of closed intervals via Euler angles etc.,} 
\end{remark}  

\section{Calculation of Certain Quantum Information Measures}
With explicit parametrizations of
purifications of a single mixed state and joint purifications of a pair of
mixed states, it is possible to compute, either in closed form or via
optimization over well defined
quantities, several quantum information measures. Many such measures are often
posed as optimization of scalar quantities over some pure states. Below we
will give two examples where this can be done in closed form. The first
is the maximal singlet fraction. A formula for this essentially appears in
\cite{hordecki}, where a full proof is not given. Further, the arguments
involved in \cite{hordecki} consist of reducing the $\delta$ matrix of
some mixed states into a normal form, which seems unmotivated.  
The proof below shows why that normal
form naturally arises. Thus, the argument provided here may be
seen as a complement to that in \cite{hordecki}. Secondly, we will
compute the joint purification closest to a given ( impure) mixed state
with the same partial traces. In general, this leads to a variation of
the Procustes problem and this variation will be formulated as the 
solution of a concrete optimization problem over the interval $[0, 2\pi]$.
For the special case when this mixed state is the
product state $\rho_{\beta}\otimes\rho_{\gamma}$ it turns out that all joint
purifications are at the same distance.
We then explain this from the perspective
of the functional being optimized in this generalized Procustes problem.

\begin{remark}
{\rm While optimization over $SO(3, R)$ or $SO(2, R)$ 
may be viewed as constrained
optimization problems and thus amenable to Lagrange mutiplier techniques,
the methods used below avoid this. In part due to the nature of the
function(al)s being optimized, it seems much better to use appropriate
parametrizations of these groups and pass directly to an unconstrained
optimization, than add further equations via the Lagrange multiplier method.}
\end{remark} 

\begin{proposition}(see \cite{hordecki})  
{\rm Consider a mixed state, $\rho$ in $C^{2}\otimes C^{2}$ represented in the
form given by Equation (\ref{intwod}). Denote the corresponding
$\delta$ matrix by $\delta_{\rho}$. Then its
maximal singlet fraction 
is given by $\frac{1}{4} ( 1 + \sum_{i=1}^{3}\sigma_{i} )$, if 
${\mbox det} \ (\delta_{\rho} ) < 0$ and by $\frac{1}{4} ( 1 + \sigma_{1}
+\sigma_{2} - \sigma_{3} )$ if ${\mbox det} \ (\delta_{\rho} ) \geq 0$   
Here, the $\sigma_{i}$ are the singular values of $\delta_{\rho}$, with
$\sigma_{i} \geq \sigma_{j}, {\mbox if} \ i < j, i,j = 1,2,3$.}  
\end{proposition}.        

\noindent Proof: The maximal singlet fraction $f(\rho )$ is defined
via the equation $f(\rho ) = {\mbox max} <\psi \mid \rho \mid \psi >$,
where the maximization is over all pure states $\psi$ 
which are maximally entangled. This collection of states is precisely the
set of pure states locally equivalent to the Bell state. It is easy to
see that this is precisely the collection of pure states, which when
represented in form Equation (\ref{intwod}) have $\beta = \gamma = 0$,
$\delta = D, D\in O(3, R), {\mbox det}\ (D) = -1$.
Now the following is true  
\[
<\psi \mid \rho \mid \psi > = {\mbox Tr}\  (\rho\rho_{\psi})
\]
So finding $f(\rho )$ amounts to maximizing ${\mbox Tr}\  (\rho\rho_{\psi})$
over all $\rho_{\psi}$ which, when represented via Equation (\ref{intwod}),
verify $\beta = \gamma = 0, \delta = D, D\in O(3, R), {\mbox det}\ (D) = -1$.

Now by Equation (\ref{producttr}), this is the same as maximizing
the quantity $\frac{1}{4} (1 + {\mbox Tr} \ (\delta_{\rho}^{T}D)$,
over $D\in O(3, R), {\mbox det} \ (D) = -1$. Equivalently it is the
maximization of ${\mbox Tr} \ [(-\delta_{\rho}^{T})V], V\in SO(3, R)$.
This is closely related to the key step in the solution of the Procustes
problem, \cite{horn}. Some of the steps require careful modification,
since the solution in \cite{horn} uses optimization over the unitary group.
In particular, the argument in \cite{horn} will not directly apply to
the case ${\mbox det}\ (\delta_{\rho}) \geq 0$. Therefore, only this
situation is addressed here. If ${\mbox det}\ (\delta_{\rho}) \geq 0$,
then ${\mbox det}\ (-\delta_{\rho}^{T})\leq 0$. Therefore, there exist
$S,T\in SO(3, R)$ such that
\[
S(-\delta_{\rho}^{T})T = - {\mbox diag}\ (\sigma_{1}, \sigma_{2}, \sigma_{3})
\]
Hence we get,
\[
{\mbox Tr} \ [(-\delta_{\rho}^{T})V]
= {\mbox Tr} \ [-S^{T} {\mbox diag}\ (\sigma_{1}, \sigma_{2}, \sigma_{3})
T^{T}V]
\]
which equals
\[
{\mbox Tr} \ [-{\mbox diag}\ (\sigma_{1}, \sigma_{2}, \sigma_{3})T^{T}VS^{T}]
\]
So this last quantity has to be maximized over $V\in SO(3, R)$.
Quite clearly the maximum occurs when $T^{T}VS^{T} = {\mbox diag}\ (-1, -1, 1)$.
Now $T, S, {\mbox diag}\ (-1, -1, 1)$ are all in $SO(3, R)$, so such a $V$
in $SO(3, R)$ always exists and is unique.
This then gives the stated expression for $f(\rho )$. 
Note $-I_{3}$ is not in $SO(3, R)$. So $f(\rho )$ cannot be increased further.

\vspace*{1.8mm}

Next we look at the distance of the joint purifications of two 
mixed states given by Bloch vectors $\beta , \gamma$ to a given impure
density matrix, $\rho_{\beta , \gamma}$ in $C^{2}\otimes C^{2}$
whose partial trace is also precisely 
the states represented by $\beta$, $\sigma$. The choice of distance is
the Hilbert-Schmidt distance $d (\rho , \rho_{\beta , \gamma })^{2}
= {\mbox Tr}\ (\rho - \rho_{\beta , \gamma })^{2}$.
Suppose the $\delta$ matrix of 
of $\rho_{\beta , \gamma })$ is denoted $E$. Then from
Equation (\ref{producttr}) it follows
\[
d (\rho , \rho_{\beta , \gamma })^{2} = \frac{1}{4}( 1  +
{\mbox Tr}\ (\delta -E)(\delta - E)^{T})
\]
Fixing one choice, $D$ for $\delta$ it follows from Th 2, that the
most general such $\delta$ is given by $DC$, with $C$ in $SO(3, R)$
satisfying $C\gamma = \gamma$. So the above quantity becomes 
${\mbox Tr} (DD^{T} + EE^{T}  - DCE^{T} - EC^{T}D^{T})$. 
Since $D, E$ are fixed minimizing this quantity is the same as maximizing
${\mbox Tr} (DCE^{T} + EC^{T}D^{T})$. But, $EC^{T}D^{T} = (DCE^{T})^{T}$.
Hence, using the cyclic invariance of trace,
the problem reduces to maximizing, over all $C\in SO(3, R)$ satisfying
$C\gamma = \gamma$, the function
\begin{equation}
\label{function}
F_{D, E}(C) = {\mbox Tr}\ (E^{T}DC)
\end{equation}

We begin with a simple observation
\begin{lemma}
{\rm Suppose $\rho_{\beta , \gamma }$ is the tensor product of the mixed
states corresponding to $\beta$ and $\gamma$, then
the function
$F_{D, E}(C)$ is constant and equals $\mid\mid\beta\mid\mid^{2}$, i.e., every
joint purification is equidistant from $\rho_{\beta , \gamma }$.}  
\end{lemma}

Proof: Since $E = \beta\gamma^{T}$, in this case, it follows
$F_{D, E}(C) = {\mbox Tr}\ (E^{T}DC) = {\mbox Tr}\ (DCE^{T})
= {\mbox Tr}\ (D\gamma\beta^{T})$, since $C\gamma = \gamma$.
Further, $D\gamma = \beta$ for pure states. So this reduces to
${\mbox Tr} \ (\beta\beta^{T}) = \mid\mid\beta\mid\mid^{2}$.

We now address the general situation. From this analysis a geometric
interpretation for the previous lemma will emerge. Now maximizing
$F_{D, E}(C)$ would reduce to the key step in the usual Procustes problem,
but for the restriction that $C\gamma = \gamma$. For $\gamma = 0$ this
is no restriction, though as in the calculation of the maximal singlet
fraction care has to be taken since the optimization is over $SO(3, R)$.
Therefore, we will study only the $\gamma\neq 0$ situation.

\begin{proposition}
{\rm The maximum of $F_{D,E}(C)$ is given by the maximum of a differentiable
function $f(\theta )$ (defined in Equation (\ref{functionthet}) below) over
the interval $[0, 2\pi]$. Thus, this maximum exists and can be found
by comparing the values of $f$ at $\theta = 0$ and at the critical
points of $f$ in $(0, 2\pi)$.} 
\end{proposition}

Proof: By the singular value decomposition
\[
E^{T}D = V\Sigma W^{T}
\]
where $V$ is a real matrix with columns $v_{i}$
eigenvectors of $E^{T}D (E^{T}D)^{T}$, $W$ is a real matrix whose columns,
$w_{i}$ are the eigenvectors of $(E^{T}D)^{T}E^{T}D$ and $\Sigma$ is a diagonal
matrix ${\mbox diag} (\sigma_{1}, \sigma_{2}, \sigma_{3})$ where the
$\sigma_{i}$ are the singular values of $E^{T}D$.

Once again by the cyclic invariance of ${\mbox Tr}$ it follows
\[
F_{D,E}(C) = {\mbox Tr}\ (\Sigma W^{T}CV)
\]
To find the diagonal entries of $W^{T}CV$, we expand $v_{i}$, $w_{i}$
in some orthogonal basis, whose first member is
member is $\gamma$ 
Denoting the components
of $v_{i}$, $w_{i}$ in the first direction by $p_{i}, s_{i}$ respectively,
we find 
\[
Cv_{i} = p_{i}\gamma + R_{\theta}x_{i}, i=1, \ldots , 3
\]
Here $p_{i} =  <v_{i}, \gamma >$, while
the $x_{i}$ are precisely the orthogonal projections of the $v_{i}$ onto
the plane perpendicular to $\gamma$ (and thus,$x_{i}$ is uniquely determined
by the $v_{i}$). In fact, since $\gamma$ is orthogonal to this plane,
$x_{i} = v_{i} - p_{i}\gamma$.
$R_{\theta}$ is the rotation through $\theta$ that the
matrix $C$ performs.

Denoting by $y_{i}, i=1,\ldots , 3$ the orthogonal projections of the $w_{i}$
onto the plane perpendicular to $\gamma$ and $s_{i} = 
<w_{i}, \gamma >$, we find 
\begin{equation}
\label{functionthet}
F_{D, E}(C) = f(\theta ) = \sum_{i=1}^{3}(\sigma_{i}p_{i}s_{i}  +
<y_{i}, \sigma_{i}R_{\theta}x_{i}>, \theta \in [0, 2\pi]
\end{equation}
So maximizing this function will yield the distance of $\rho_{\beta , \sigma}$
from the set of joint purifications of the mixed states represented by
the Bloch vectors $\beta , \gamma$.

\begin{remark}
{\rm The situation covered by Lemma can now be explained as follows.
$E = \beta\gamma^{T}$. So the singular values of $E^{T}D$ are the
eigenvalues of $E^{T}DD^{T}E$. This is the rank one matrix 
$\mid\mid\gamma\mid\mid^{2}\gamma\gamma^{T}$. Hence the singular
values are $(\mid\mid\gamma\mid\mid^{2} , 0, 0)$. Further, $v_{1} =
w_{1} =  \gamma$.
Hence, $x_{1} = 0$. So $F_{D, E}(C)$ us independent of $\theta$, i.e.,
of $C$. Finally, since $\mid\mid\gamma\mid\mid = \mid\mid\beta\mid\mid$,
the stated value for $F_{D, E}(C)$ is indeed obtained.} 
\end{remark}.

\begin{remark}
{\rm In Proposition 3.1, the optimization over $SO(3, R)$ did not require
more than elementary aspects of the group structure of $SO(3, R)$. 
For related optimization problems we found the following parametrization
of $SO(3, R)$ useful. Let the normalized eigenvector of the generic
$D\in SO(3, R)$, belonging to the eigenvalue $1$,
be written in spherical coordinates in the form 
$ e_{D} =
(\cos\theta_{D}\sin\phi_{D}, \sin\theta_{D}\sin\phi_{D} , \cos\phi_{D})$.
Denote the angle of rotation $D$ performs (counterclockwise) in the plane
orthogonal to this eigenvector be $\psi_{D}$. Then the effect of $D$
on any vector in $v\in R^{3}$ is $<v, e_{D}>e_{D} + R_{\psi_{D}}
(v - <v, e_{D}>e_{D})$. Since the typical optimization problem arising in
contexts similar to the ones in this work involve inner products and norms,
this representation of $SO(3, R)$ seems superior to others 
(over Euler angles or Givens rotations, for instance). In this representation
the optimization reduces to optimizing a function $F(\phi , \theta , \psi )$
over $[0, \pi]\times [0, 2\pi]\times [0, 2\pi]$.}
\end{remark}

\section{Conclusions}
This note yielded a complete parametrization of purifications and joint
purifications of $2\times 2$ density matrices. This enabled a reformulation
of the calculation of quantum information measures as optimization problems
over the real orthogonal groups. In particular, a solution to one variation
of the classical Procustes problem was provided. It would be interesting
to extend this to density matrices in higher dimensions. The first ingredient
in this a clear description of density matrices and pure states, going
beyond the fact that they are expressible as real linear combination of
certain matrices and necessarily have trace 1. In other words, a full
characterization of this real vector of coefficients in this expression
is desirable. This is partially addressed by the appendix.

\section{Acknowledgements}
This work is supported, in part, by a NSF grant, DMS - 0072415.

\section{ Appendix}
How may one generalize this explicit parametrization of purifications? 
The first ingredient is a Bloch sphere like picture of density matrices
in the appropriate $C^{n}$. For this there are two standard points of 
departure beyond 
the setting here. One is to consider higher tensor products of
two dimensional spaces or to consider twofold tensor products of spaces
of dimension higher than two. In an attempt to achieve this for the second
route, we consider the question of describing, in a Bloch sphere fashion,
density matrices of a single system first. This already is
quite an arduous job as may be seen below.

First we represent a typical $n\times n$ density matrix in the following
form:
\begin{equation}
\label{innd}
\rho = \frac{1}{n}(I_{n} + \sum_{i=1}^{n^{2}-1}\beta_{i}\lambda_{i}) 
\end{equation}
where the matrices $\lambda_{i}$ satsify i) $\{i\lambda_{k}, k=1, \ldots ,
n^{2}-1 \}$ is an orthogonal basis for $su(n)$; ii) their ``Jordan"
commutator satisfies, $(\lambda_{k}\lambda_{l} + \lambda_{l}\lambda_{k})
= \frac{4}{n}I_{n} + \sum_{i=1}^{n^{2}-1}d_{kli}\lambda_{i}$, with the
$d_{kli}$ a symmetric tensor. Such bases always exist. In principle,
the $d_{kli}$ can be found, though we are aware of their enumeration
only for $n=3,4$ for one particular choice of such a basis, \cite{muller}.
For $n=3$ this basis is precisely the set of the Gell-Mann matrices.    
The reasons for choosing
this representation is twofold i) eventually we wish to take partial traces of
density matrices in $C^{n}\otimes C^{n}$. Therefore, having a basis for
density matrices in $C^{n}$ which is
maximally traceless will facilitate this computation.
ii) The $\lambda_{i}$'s properties are well known.

For any $n$, using the symmetric tensor, $d_{ijk}$, define a new vector
$x\cup y \in R^{n^{2} - 1}$
starting with two vectors $x, y\in R^{n^{2} -1}$ via
\[
x\cup y = (\sum_{j,k = 1}^{n^{2} - 1} d_{1jk}x_{j}y_{k}, \ldots , 
\sum_{j,k = 1}^{n^{2} - 1} d_{ijk}x_{j}y_{k}, \ldots )
\]

The goal now is to characterize the form of the vector $\beta\in R^{n^{2}-1}$
which ensures $\rho$ is a density matrix. Since $\rho$ is the square
of a Hermitian matrix, $M$, we take a generic such $M$, square it and insist
its trace be $1$. This leads to the following characterization of
mixed states and in particular, pure states.

\begin{proposition}
{\rm Every density matrix can be represented in the form in Equation
(\ref{innd}) with
$\beta = \frac{2\kappa}{n}\beta_{0} + \frac{\beta_{0}\cup\beta_{0}}
{n}$, where $\beta_{0}$ is any vector in $R^{n^{2}-1}$ with 
$\mid\mid\beta_{0}\mid\mid^{2}\leq \frac{n^{2}}{2}$ and $\kappa =
+ \sqrt{\frac{n^{2}- 2 \mid\mid\beta_{0}\mid\mid^{2}}{n}}$.
Conversely any Hermitian matrix admitting such a representation is
necessarily a density matrix. $\rho$ is pure precisely
if it can be represented in the form in Equation \ref{innd})  
with $<\beta , \beta > = \frac{n^{2}-n}{2}$ and $(n-2)\beta =
\beta \cup \beta$.}
\end{proposition}

Note that when $n=2$, this is precisely the usual Bloch sphere,
since the Pauli matrices anti-commute, i.e. $\beta \cup \beta = 0$ if
$n$ is 2.
For $n=3$, states are pure if the vector $\beta \in R^{8}$ is of
Euclidean length $\sqrt{3}$ and $ \beta = \beta \cup \beta$.
One can now use this to write down a Bloch sphere like picture for
density matrices in $C^{n}\otimes C^{n}$. This characterization involves both
the Lie product and the Jordan product.  However, since even for
single systems, the picture provided above requires further analysis
the details will be considered in a future study.
Indeed even analyzing the 
implications of an equation of the type $ \beta = \beta \cup \beta$
for density matrices in $C^{3}$ requires
further work. However, given specific Hermitian operators (for one or
two particles) one can use this
characterization to check if they are indeed density 
matrices and even if they are pure.

\end{document}